\documentclass[12pt,preprint]{aastex}
\title {Can Early Type Galaxies Evolve from Fading the Disks of Late Types?}
\author {Daniel Christlein and Ann I. Zabludoff}
\affil {Steward Observatory, The University of Arizona\\933 N Cherry Ave, Tucson 85721 AZ}
\email {dchristlein@as.arizona.edu \\ azabludoff@as.arizona.edu}
\begin {document}
%\begin {titlepage}
\shortauthors{Christlein \& Zabludoff}
\shorttitle{Disk Fading} 
\begin {abstract}
We examine whether early-type galaxies in clusters may have evolved from later types by the fading of their disks (e.g., as a result of ram-pressure stripping or strangulation) or by enhancement of the bulge luminosity (e.g., due to tidal interactions and mergers). For this purpose, we compare the bulge and disk luminosities of early- and late-type galaxies and of galaxies at different radial distances from the cluster center. We find that, in order for early-type galaxies, including S0s, to have evolved from late-type galaxies, their bulge luminosities must have been physically enhanced. Disk fading models cannot explain the differences observed. We then show that galaxy bulges are systematically brighter at small projected distances from the cluster center, while disk luminosities are uncorrelated with cluster-centric distance. Our results suggest that bulge enhancement, not disk fading, distinguishes early from late types and is thus at least partially responsible for the morphology-environment relation of bright cluster galaxies.
 \end{abstract}
\keywords{galaxies: clusters: general --- galaxies: evolution --- galaxies: luminosity function --- galaxies: structure}
%\end{titlepage}

\section{INTRODUCTION}

Galaxies have a wide range of morphologies, and the mechanisms that are responsible for generating this diversity are not completely understood. Elliptical galaxies may be the result of major mergers \citep{bh92}. Spirals may represent a more pristine population undisturbed by major disruptive events. The origin of intermediate types --- particularly S0s --- has been the subject of much debate.

One proposed scenario to explain the apparent increase in the fraction of S0s in rich clusters over time \citep{dressler96} is that, as spirals enter dense environments, their reservoirs of neutral gas are cut off either by disrupting further accretion (i.e., by strangulation; Larson, Tinsley \& Caldwell 1980; Balogh, Navarro \& Morris 2000; Bekki, Couch \& Shioya 2002), or by removing gas directly from the disk (i.e., via ram-pressure stripping; Gunn \& Gott 1972). Star formation in the disk would then cease, and the disk would evolve passively and fade as high-mass stars die. Studies of clusters at higher redshift \citep{couch98,dressler99,poggianti99} have uncovered a population of recently star forming ($\sim$ 1-2 Gyr earlier), but now quiescent, galaxies with normal disk morphologies that those authors cite as possible examples of systems whose star formation was disrupted as described above.

Another possibility is that galaxy-galaxy interactions, including close tidal encounters and mergers, increase the luminosity of the bulge component by heating the central parts of the disk or triggering star formation in the center \citep{barnes,bekki98,mihos94}. Extensive observations demonstrate that such interactions do increase central star formation rates \citep{lonsdale,kennicutt87,liukennicutt} and can produce bulge-enhanced merger remnants \citep{yang}. These interactions are likely in clusters with substructure (i.e., poor groups accreted from the field), where the relative speeds and internal velocity dispersions of subcluster members are similar.

Both scenarios increase the bulge fraction, $B/T$, the fraction of the total luminosity of a galaxy associated with the bulge. The first scenario increases $B/T$ by reducing the luminosity contribution from the disk. The second scenario increases $B/T$ by increasing the luminosity of the bulge. Could either mechanism be responsible for generating the morphological sequence in clusters?

Past observational work has attempted to address this question by comparing bulge fractions or bulge luminosities to Hubble type and/or environment \citep{dressler80,boroson,simien,solanes89}, with often conflicting conclusions. \citet{dressler80} finds that the bulges of S0s are more luminous than those of spirals, and that their luminosity is increased in denser environments. In contrast, \citet{solanes89}, working from the same sample, argue that disk luminosities are decreased in dense environments. \citet{boroson} finds bulge fractions of S0s to be similar to those of spirals above a certain $B/T$, while \citet{simien} find S0s to have systematically larger bulge fractions.

Several problems affect these studies: 1) While \citet{dressler80} and \citet{solanes89} consider a large, apparent magnitude-limited sample of cluster galaxies, they are not able to spectroscopically confirm cluster members and must resort to statistical background subtraction. \citet{valotto} have pointed out that background subtraction can fail in optically selected cluster samples, because cluster selection is biased towards clusters with higher-than-average background contamination. As a consequence, artificial correlations between environment and galaxy morphology could arise from interloping galaxies, especially at large projected radii from the cluster center, if the background subtraction is not perfect. 2) Hubble types are generally used as morphological quantifiers, a measure that is subjective and not easily reproducible. 3) The bulge-disk decompositions in \citet{dressler80} (also used by Solanes et al.) are done visually and affected by larger uncertainties than achievable with the best automated mechanisms and digital imaging. Purely visual analysis techniques may also fail to resolve some bulges and disks, leading to incompletenesses that are difficult to quantify. 4) None of the studies above quantify how their morphological classifications may be influenced by the bulge fraction. This bias, whether conscious or otherwise, may dramatically affect the interpretation of those results, because it could introduce intrinsic correlations between bulge and disk luminosities and morphological type. 5) The impact of bulge enhancement and disk fading is on the luminosity distributions of the bulge and disk components, and the luminosity functions of bulges and disks are thus the best discriminators between the two evolutionary hypotheses outlined above, but only \citet{solanes89} determine a luminosity function. This determination is semi-empirical and not calculated directly from the observed luminosity distribution of bulges. More recently, \citet{benson02} have calculated bulge and disk luminosity functions directly, but their sample is too small to hold any discriminatory power over the evolutionary mechanisms described above.

In this paper, we compare the luminosity functions for bulges and disks, calculated directly from a sample of several hundred spectroscopically-confirmed cluster galaxies \citep{cz2003} to determine whether early type galaxies could have evolved from late type galaxies by disk fading alone. We use the bulge fraction, $B/T$, rather than Hubble type, as a quantitative, reproducible measure of morphology, which enables us to account for any bias that could introduce intrinsic correlations between bulge luminosity and morphology. By examining bulge and disk luminosities as a function of projected radial distance from the cluster center, we test whether the morphology-environment relation \citep{dressler80} could have been generated principally by bulge-enhancing or disk-diminishing effects. 

\section{THE DATA}

Our sample consists of cluster galaxies from a spectroscopic and $R$-band imaging survey of six nearby clusters (A85, A496, A754, A1060, A1631, A3266). The spectroscopic survey ensures that contamination of the sample by background field galaxies is minimized. Table \ref{tabclusters} lists the kinematic properties of these six clusters for $H_{0}=71$ km s$^{-1}$ Mpc$^{-1}$, $\Omega_{m}=0.3$ and $\Omega_{\Lambda}=0.7$, as applied throughout this paper. For details regarding the survey and data reduction, see \citet{cz2003}. 

\begin{deluxetable}{lrrrrcrr}
\tabletypesize{\scriptsize}
\tablecaption{The Cluster Sample}
\tablewidth{0pt}
\tablehead{
\colhead{Cluster}&\colhead{\#}&\colhead{$\overline{cz}$}&\colhead{$\Delta m$}&\colhead{$cz$ range}&\colhead{$\sigma$}&\colhead{$r_{sampling}$ } \\
\colhead{       }&\colhead{Galaxies}&\colhead{ [km/s]}&\colhead{[mag]}&\colhead{[km/s]}&\colhead{[km/s]}&\colhead{[Mpc]} }
\startdata
 A1060&252&$3683\pm46$ &33.59&2292  - 5723 &$ 724\pm31$&0.67\\
 A496 &241&$9910\pm48$ &35.78&7731  - 11728&$ 728\pm36$&1.76\\
 A1631&340&$13844\pm39$&36.53&12179 - 15909&$ 708\pm28$&2.42\\
 A754 &415&$16369\pm47$&36.90&13362 - 18942&$ 953\pm40$&2.83\\
 A85  &280&$16607\pm60$&36.94&13423 - 19737&$ 993\pm53$&2.87\\
 A3266&331&$17857\pm69$&37.10&14129 - 21460&$1255\pm58$&3.07\\
\enddata
\tablecomments{\# is the number of sampled galaxies per cluster. $\overline{cz}$ is the mean velocity, $\Delta m$ the distance modulus (for $H_{0}=71$ km s$^{-1}$ Mpc$^{-1}$). ``cz range'' is the velocity range spanned by cluster members, $\sigma$ is the line-of-sight velocity dispersion, and $r_{sampling}$ is the physical radius sampled.} 
\label{tabclusters}
\end{deluxetable}

\section{BULGE-DISK DECOMPOSITION}

We use the GIM2D software \citep{gim2d} to perform a two-dimensional decomposition of the galaxy images into a bulge component, described by a de Vaucouleurs surface brightness profile \citep{devauc}, and a disk component with an exponential surface brightness profile. Prior to the fit, we transform all galaxy images to a fiducial rest frame $cz=17858$ $km/s$ (corresponding to mean the velocity of the most distant cluster, A3266) by fading the surface brightnesses by $(z_{cosmological}+1)^{-2}(z_{total}+1)^{-2}$, smearing the images to achieve a consistent FWHM of 2 arcsec, and rebinning their pixels with the new angular diameter distance to achieve the same physical resolution per pixel. This approach ensures that determinations of $B/T$ are internally consistent among the clusters, which span a mean velocity range from 3682 to 17858 km/s. The final catalog contains bulge-disk decompositions of 1637 galaxies (1304 of them for galaxies with $M_{R}\leq-19.25$). The full catalog, including $B/T$ values, appears in a subsequent paper (Christlein \& Zabludoff 2004, in prep.).

% mordist.eps > f1.eps
\begin{figure}
\plotone{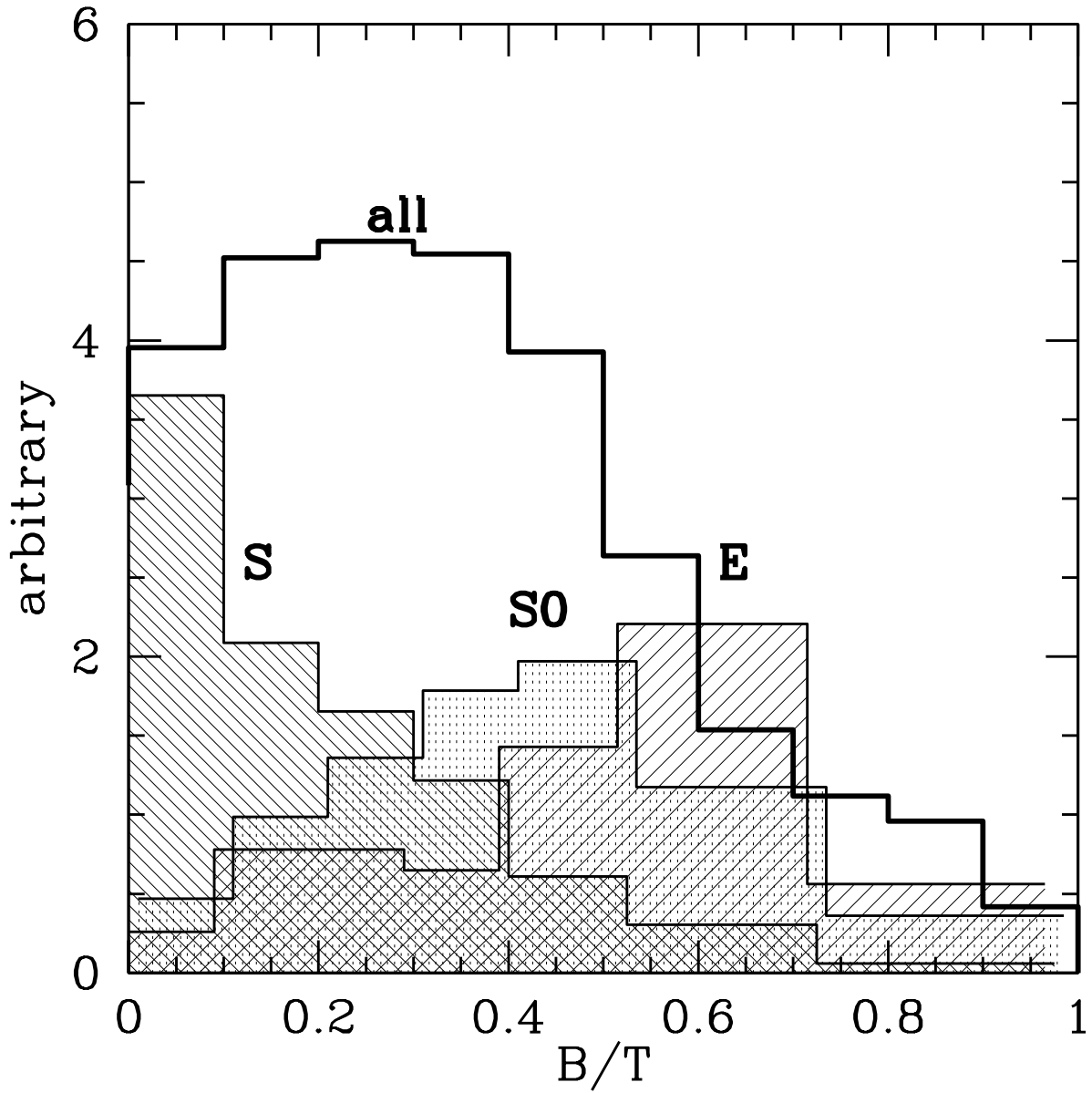}
\caption{Distribution of measured $B/T$ for galaxies identified in the NASA Extragalactic Database (NED) as spirals (S), S0s, and ellipticals (E) in units such that the area under each curve is unity. The bold, unshaded histogram shows the $B/T$ distribution for all galaxies in the sample, regardless of whether Hubble types are available for them in NED (scaled by a factor of 3.5 for better legibility).
 For display purposes, the histograms have been slightly displaced along the $B/T$ axis. Spirals have low $B/T$, with a median $B/T<0.2$. S0s and Es have very broad distributions with medians $B/T\approx0.45$ and $B/T\approx0.6$, respectively. Many galaxies identified in the literature as ellipticals have significant disk components.}
\label{fig_mordist}
\end{figure}

Fig. \ref{fig_mordist} shows the distribution of $B/T$ for galaxies that are identified in the NASA Extragalactic Database (NED) as S, S0, or E, as well as the $B/T$ distribution for all sample galaxies with and without literature classifications. Because of the inhomogeneity of the literature sources, this figure should only be regarded as an approximate orientation. The distribution of $B/T$ in our sample is roughly comparable to those of other studies such as \citet{vy}, although we find a lower fraction of extreme late-type galaxies ($B/T<0.1$) and a higher fraction of intermediate-type galaxies ($B/T\approx0.3$, typical of early-type spirals) than their sample, which consists only of field and group galaxies. Spirals are mostly confined to small $B/T$. There are very few pure bulge systems in our sample, and we find significant disk components even in galaxies classified as ellipticals (c.f., Rix \& White 1990). Such low-surface brightness features are typically more difficult to identify on photographic plates, which may explain why such systems have been classified as ellipticals. The fact that our classifications are in good agreement with galaxies typed as spirals in the literature supports this hypothesis.

To test our quantitative morphological classifications, we visually examine all literature-classified ``ellipticals'' with the lowest $B/T$ values ($B/T<0.4$). For the five most luminous among them, the reason for the low $B/T$ is usually apparent (brightest cluster galaxies with extended envelopes or interacting systems with extended features that could be disks or tidal debris). Although these ``disk'' features may be of a different nature than normal spiral disks, they are reproduced well by GIM2D, and so we do not discard them in order to avoid introducing subjective selection criteria. Most low-$B/T$ ``ellipticals'' are fainter ($M_{R}>-21$), and GIM2D also fits these systems very well. Attempts to fit these systems with pure bulge models typically result in an increase both in the reduced $\chi^{2}_{\nu}$ and in the fraction of residual light after the model has been subtracted. Overall, $B/T$ is well correlated with Hubble type for our sample (the Spearman rank correlation coefficient is $r=0.57$ at $12\sigma$ deviation from the no-correlation hypothesis) and therefore suitable for our analysis. 

\section{CALCULATING THE BULGE AND DISK LUMINOSITY FUNCTIONS}

To calculate the distribution functions of bulge and disk luminosities, we need to consider not only bulge and disk luminosities, but also total luminosity and surface brightness, because these quantities determine the completeness of the spectroscopic and morphological catalogs. This multi-variate problem is best solved with the Discrete Maximum Likelihood (DML) algorithm \citep{cmz2003}, which does not require any prior assumptions regarding the analytical form or dimensionality of the galaxy distribution function. This algorithm calculates statistical weighting factors that incorporate the completeness corrections and volume corrections for each galaxy. These corrections account statistically for the fact that the spectroscopic and morphological catalogs are not complete and that a galaxy of a given absolute magnitude may be observable in some, but not all, of the clusters in our sample. 

 The spectroscopic catalog is the major source of incompleteness; depending on the surface brightness and cluster, the completeness of the catalog for $m_{R}=18$ galaxies is in the range $\sim0.25$ to $\sim0.5$. The morphological catalog is mostly complete; the fraction of galaxies in the spectroscopic catalog that have bulge-disk decompositions is never smaller than $\sim85\%$ for any $m_{R}<18$.

We use the weighting factors calculated by the DML to reconstruct the bulge and disk luminosity functions (LFs, \S \ref{bdlfs}) and to calculate correlation coefficients for the bulge and disk luminosity-environment relations (\S \ref{bdenv}). Even for the faintest galaxies that we analyze in this paper ($M_{R}=-19.25$), the final weighting factor (incorporating completeness and volume corrections as determined by the maximum likelihood method) is never larger than $\sim 2.5\times$.

We propagate uncertainties in $B/T$ into the final bulge and disk luminosity functions with a Monte Carlo algorithm. For each of 100 realizations, the algorithm draws a value of $B/T$ from the uncertainty interval for each galaxy. We then determine the mean and scatter of the value of the luminosity function in each magnitude bin due to these $B/T$ uncertainties and add the scatter in quadrature to the Poisson errors. This is a conservative procedure, because the $B/T$ uncertainty intervals are given at the $99\%$ confidence level by GIM2D, and because errors in different magnitude bins due to $B/T$ uncertainties are really correlated for any given realization (i.e., if a galaxy is presumed to be in one bin for a given choice of $B/T$, it cannot lie in any other bin). 

\section{RESULTS AND DISCUSSION}

\subsection{Bulge- and Disk Luminosity as a Function of Morphology}
\label{bdlfs}

To isolate the mechanism by which early-type galaxies may evolve from late types, we compare the bulge and disk luminosities of galaxies with different morphologies. We split the sample into six independent subsamples, selected by $B/T$. Table \ref{samptab} shows the $B/T$ intervals for each sample, the median $B/T$, and the number of galaxies in each bin. The next three columns show the percentages of all S, S0, and E-type galaxies that fall into each bin, based on literature classifications from NED. Because of the inhomogeneity of the literature sources and the problems discussed earlier that are associated with classifications based on photographic plates, these percentages should only be taken as a rough calibration of the distribution of galaxies in $B/T$. 

Our analysis does not extend to deep magnitudes: in all cases, $M_{R}<-19.25$, the absolute magnitude limit of the spectroscopic and morphological catalogs for the most distant cluster, A3266. We therefore focus our work on the bright end of the luminosity function and characterize it by fitting Schechter functions \citep{schechter76} to the bulge and disk luminosity functions in each subsample. We impose the constraint that the faint end slope is flat ($\alpha=-1$). By fixing $\alpha$ at a constant value, the Schechter parameter $M^{*}$ becomes a direct measure of the characteristic magnitude of the bright end exponential cutoff. We refer to the $M^{*}_{R}$ obtained with this slope constraint as $M^{*}_{R}(\alpha=-1)$.

\begin{deluxetable}{lrrrrrrrrr}
\tabletypesize{\scriptsize}
\tablecaption{Morphological Subsamples and Schechter Fits }
\tablewidth{0pt}
\tablehead{
\colhead{Subsample} & \colhead{B/T$_{min}$} & \colhead{B/T$_{max}$} & \colhead{B/T$_{med}$} & \colhead{\#} & \colhead{\% S} & \colhead{\% S0} & \colhead{\% E} & \colhead{$M^{*}_{bulge}$} & \colhead{$M^{*}_{disk}$}
\\
\colhead{} & \colhead{ } & \colhead{} & \colhead{} & \colhead{Galaxies} & \colhead{               } & \colhead{               } & \colhead{               } & \colhead{$(\alpha=-1)$} & \colhead{$(\alpha=-1)$} }
\startdata
A & 0.0 & 0.19 & 0.05 & 498$^{+31}_{-4}$ & 57.4 & 14.6 & 10.4 & $ -19.32_{-0.37}^{+0.39} $&$ -20.74_{-0.11}^{+0.12}$\\
B & 0.2 & 0.29 & 0.25 & 247$^{+6}_{-28}$ & 16.5 & 13.6 & 7.8 &  $ -19.83_{-0.21}^{+0.23} $&$ -20.99_{-0.16}^{+0.18}$\\
C & 0.3 & 0.39 & 0.35 & 239$^{+11}_{-11}$ & 12.2 & 17.8 & 6.5 & $ -20.04_{-0.22}^{+0.22} $&$ -20.84_{-0.17}^{+0.17}$\\
D & 0.4 & 0.49 & 0.45 & 239$^{+5}_{-21}$ & 6.1 & 19.7 & 14.3 &  $ -20.91_{-0.21}^{+0.21} $&$ -21.13_{-0.14}^{+0.16}$\\
E & 0.5 & 0.69 & 0.59 & 310$^{+19}_{-22}$ & 6.1 & 23.5 & 44.1 & $ -21.62_{-0.12}^{+0.13} $&$ -21.21_{-0.12}^{+0.13}$\\
F & 0.7 & 1.0 & 0.91 & 185$^{+36}_{-0}$ & 1.7 & 10.8 & 16.9 &   $ -21.23_{-0.13}^{+0.14} $&$ -20.00_{-0.35}^{+0.39}$\\
\label{samptab}
\enddata
\end{deluxetable}

\label{selection}

\begin{figure}
\plotone{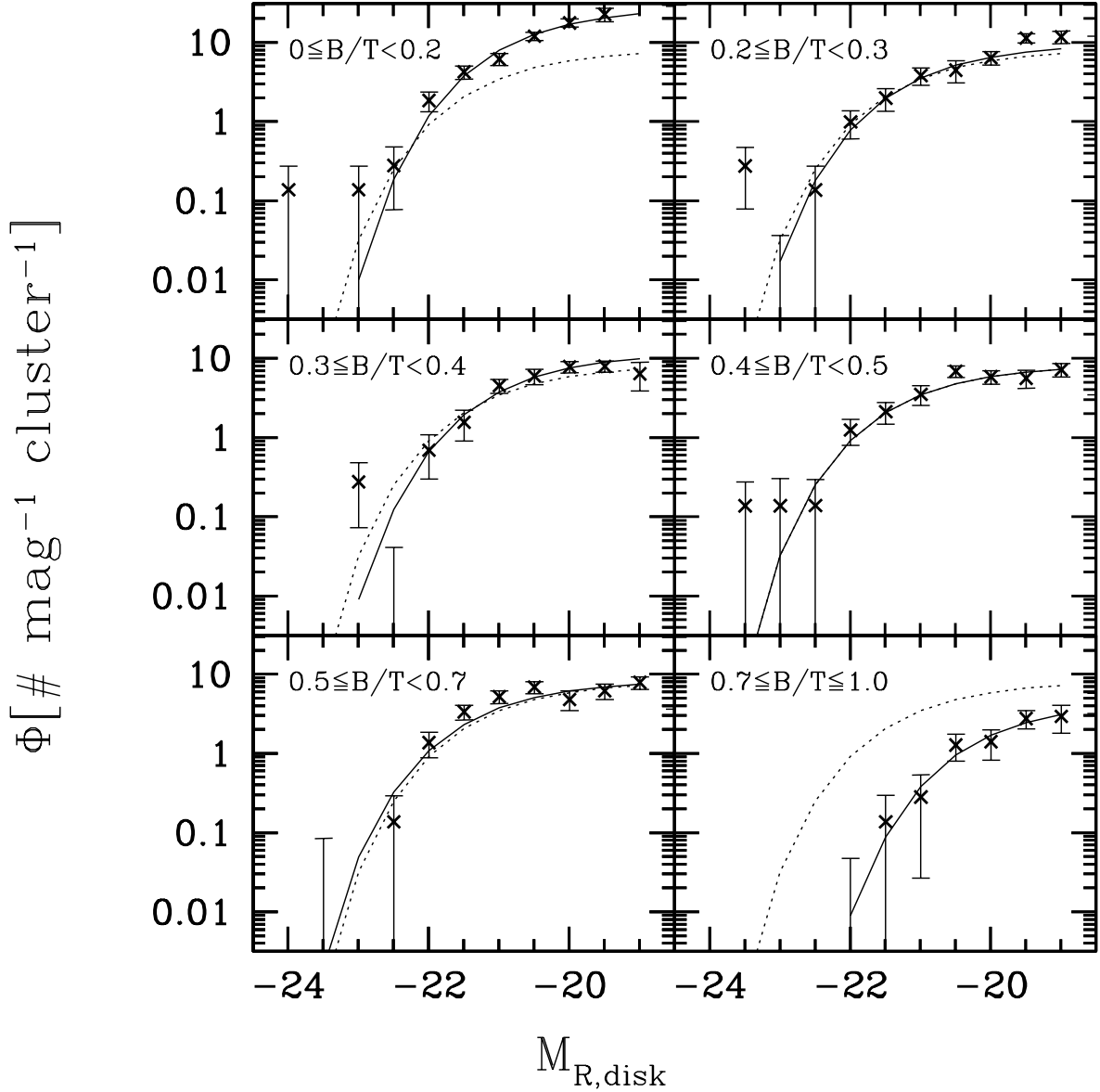}
\caption{Luminosity functions and Schechter fits (with $\alpha=-1$) for disks for six subsamples selected by their bulge fraction, B/T. See text and Table \ref{samptab} for the definition of the subsamples. For orientation, the Schechter fit for an intermediate subsample ($0.4\leq B/T<0.5$) has been marked in each panel. The bright end of the disk LF shows little variation along the morphological sequence (with the exception of the disk LF of the earliest-type galaxies, which is significantly fainter). }
\label{ff1d}
\end{figure}

\begin{figure}
\plotone{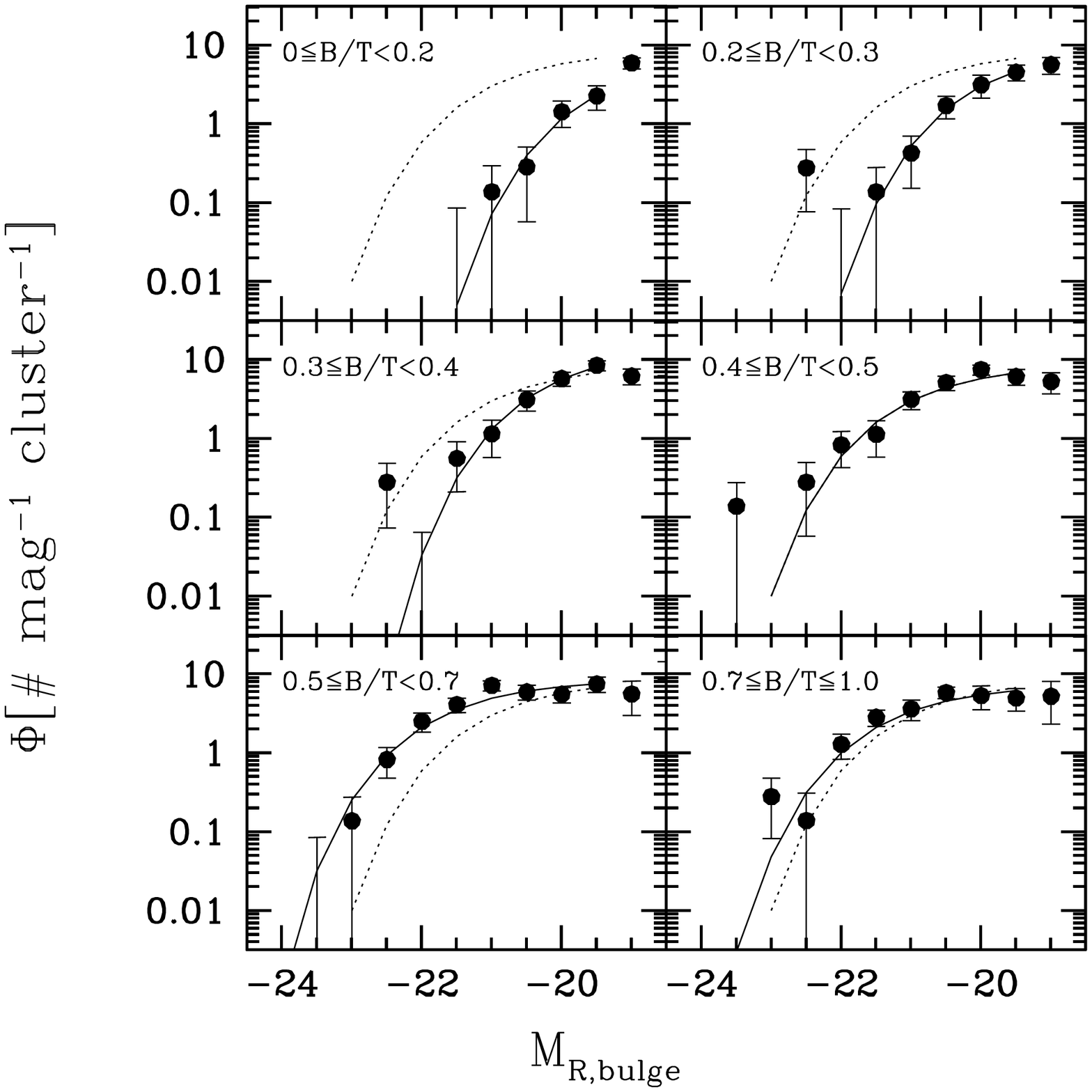}
\caption{Luminosity functions and Schechter fits (with $\alpha=-1$) for bulges for six subsamples selected by their bulge fraction, B/T. See text and Table \ref{samptab} for the definition of the subsamples. For orientation, the Schechter fit for an intermediate subsample ($0.4\leq B/T<0.5$) has been marked in each panel. From the late to early types, the bright end of the bulge LF grows more luminous by $\sim 2$ mag.}
\label{ff1b}
\end{figure}

Figs. \ref{ff1d} and \ref{ff1b} show all twelve LFs (six bulge LFs and six disk LFs) and the Schechter fits with $\alpha=-1$. With the exception of the disk LF for $0.5\leq B/T<0.7$, all fits are consistent with the LFs within $2\sigma$. Fig. \ref{ff1d} shows that the bright ends of the disk LFs are similar, except for the earliest type galaxies ($B/T\leq0.7$), whose LF is considerably fainter. In contrast to the disk LFs, the bright ends of the bulge LFs (Fig. \ref{ff1b}) spread out over $\sim 2$ magnitudes from the latest- to the earliest-type subsample.

%ff1.paneld.eps > f2.eps

%ff1.panelb.eps > f3.eps

\begin{figure}
\plotone{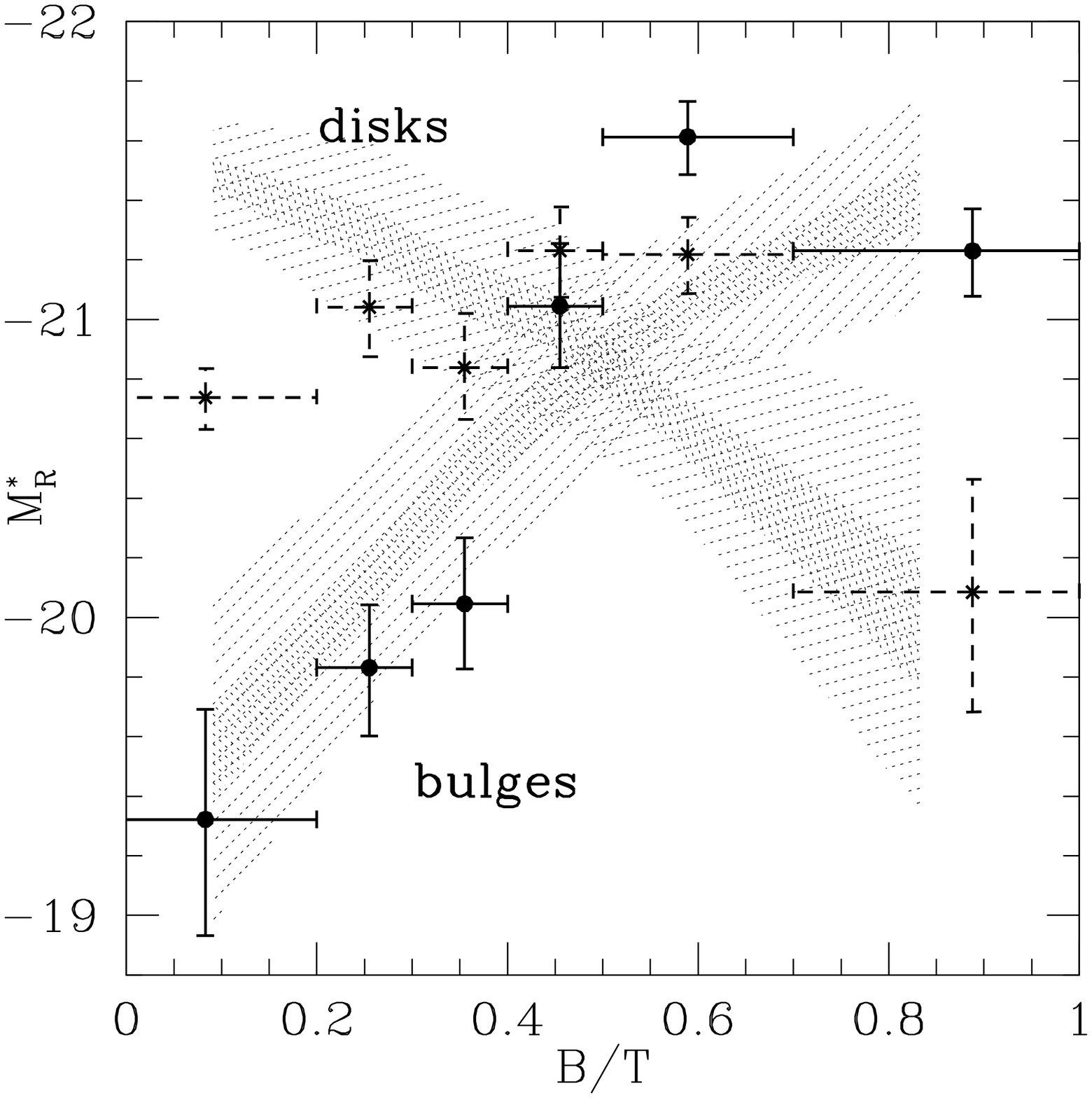}
\caption{Characteristic bright magnitude, $M^{*}_{R}(\alpha=-1)$, as a function of $B/T$ for bulge (circles) and disk luminosity functions (crosses). Error bars in $M^{*}_{R}$ show $1\sigma$ uncertainties from the Schechter fits. Error bars in $B/T$ indicate bin width. Shaded areas show the $1\sigma$ and $2\sigma$ uncertainty intervals from the ``B/T bias'' Monte Carlo sample discussed in the text. Over most of the morphological sequence, the correlations are more negative than expected, favoring bulge enhancement over disk fading as the process that may transform late-type into early-type galaxies.}
\label{f2}
\end{figure}

 The last two columns of Table \ref{samptab} list the values of $M^{*}_{R}(\alpha=-1)$ for bulges and disks in all six subsamples, and Fig. \ref{f2} shows this information as a function of $B/T$. 

%f1.eps > f4.eps

In Fig. \ref{f2}, $M^{*}_{bulge}(\alpha=-1)$ is negatively correlated with $B/T$ over a wide range of $B/T$. $M^{*}_{disk}(\alpha=-1)$ shows a weaker, but also negative, correlation with morphology. Note that the two slopes are not statistically independent, because, for any given $B/T$, $M_{bulge}-M_{disk}$ is fixed. Therefore, the only free parameter that places constraints on the evolution mechanisms of these galaxies is the absolute value of either slope.

Plotting a quantity $M^{*}_{disk}(\alpha=-1)$ or $M^{*}_{bulge}(\alpha=-1)$ that is explicitly dependent on $B/T$ versus $B/T$ introduces a bias, which we refer to as the ``B/T bias''. In selecting high-$B/T$ galaxies to determine $M^{*}_{bulge}(\alpha=-1)$, we are typically selecting galaxies with brighter bulges, but also galaxies with fainter disks, which creates an intrinsic correlation between $B/T$ and the luminosities of bulges and disks at a given $B/T$. This effect depends on parameters such as the bin size and the sample's $B/T$ distribution, and cannot be quantified analytically. To account for this bias and determine if the trends in Fig. \ref{f2} favor disk fading or bulge enhancement, we use a Monte Carlo analysis to calculate the range of expected slopes of the $M^{*}_{bulge}(\alpha=-1)$-$B/T$ and $M^{*}_{disk}(\alpha=-1)$-$B/T$ relations under the following null hypotheses: 1) B/T bias is solely responsible for the observed trends; 2) disk fading occurs in addition to the effects of B/T bias; 3) bulge enhancement operates in addition to the bias. We construct the Monte Carlo samples for these null hypotheses in the following ways:

(a) ``B/T Bias'': Can B/T bias alone explain the observed trends? In this case, the observed trends in $M^{*}_{disk}$ and $M^{*}_{bulge}$ with $B/T$ should arise even if the $B/T$ values are uncorrelated with the total absolute magnitude $M_{T}$. The most straightforward test of this hypothesis is therefore to scramble the observed $B/T$ values with respect to $M_{T}$, using them to calculate a new $M_{bulge}$ and $M_{disk}$ for each galaxy. B/T bias is then the only source of correlations between $M_{bulge}$ and $B/T$ and between $M_{disk}$ and $B/T$.

We construct the Monte Carlo sample by associating random values of $M_{T}$ from the completeness-corrected galaxy luminosity function with values of $B/T$ drawn from the non-completeness-corrected $B/T$ distribution of the sample. This ensures that the progenitor population is drawn from a realistic luminosity function, but that the $B/T$ distribution of the Monte Carlo sample is the same as in the observed sample.

For each Monte Carlo sample (of several thousand), we then calculate LFs and Schechter functions for the bulge and disk components. We compare the trends of $M_{disk}^{*}(\alpha=-1)$ and $M_{bulge}^{*}(\alpha=-1)$ with $B/T$ that we recover from the Monte Carlo samples to the observed trends to decide whether B/T bias can account for the latter.

The shaded regions in Fig. \ref{f2} show the range of $M^{*}_{disk}$ and $M^{*}_{bulge}$ recovered from this null hypothesis in each $B/T$ bin. Even from visual inspection, it is clear that the observed slopes are more negative than predicted by the B/T bias hypothesis.

(b) ``Disk Fading'': With this null hypothesis, we examine whether a disk fading mechanism, in combination with B/T bias, can reproduce the observed trends in $M^{*}_{disk}$ and $M^{*}_{bulge}$ with $B/T$. Under the Disk Fading null hypothesis, any galaxy with a $B/T$ characteristic of early-type galaxies has been generated from a progenitor population of low-$B/T$ galaxies by a reduction in the disk luminosity. 

We incorporate the disk fading mechanism into the Monte Carlo catalog in the following way: we begin by scrambling all $B/T$ values in the catalog with respect to $M_{T}$, as we did for the ``B/T Bias'' hypothesis. Then, for each mock galaxy with $B/T>0.3$, we calculate the amount of disk fading necessary to generate it from a progenitor with $B/T=0.3$, and apply it to the disk and total luminosity of the galaxy. As we did for the ``B/T bias'' hypothesis, we then calculate $M^{*}_{disk}$ and $M^{*}_{bulge}$ again in each B/T bin. 

In the absence of any bias, this hypothesis would generate a positive correlation between $M_{disk}^{*}$ and $B/T$ (i.e., galaxies with larger $B/T$ have fainter disks), while it would not introduce a correlation between $M_{bulge}^{*}$ and $B/T$.

Our choice of $B/T\approx0.3$ for the hypothetical progenitor population is a conservative one, because 1) $73\%$ of spirals have $B/T<0.3$, and 2) the peak of the spiral distribution is at $B/T<0.2$, while the peak of the S0 distribution is at $B/T\approx0.45$. If the progenitors of today's cluster galaxies were high-redshift field galaxies, it is likely that they had even lower $B/T$ values than today's population of cluster spirals.

(c) ``Bulge Enhancement'': Under this hypothesis, galaxies with high $B/T$ are generated from low-$B/T$ galaxies by an increase in the bulge luminosity. We construct the Monte Carlo samples for this hypothesis analogously to the ``Disk Fading'' hypothesis, except that, instead of fading the disks of our Monte Carlo sample members to adjust $B/T$, we increase the luminosity of the bulges. 

Without B/T bias, this effect would lead us to expect a negative correlation between $M_{bulge}^{*}$ and $B/T$, and no correlation between $M_{disk}^{*}$ and $B/T$.\footnote{The predictions of these three null hypotheses could also be described in terms of a relation between total galaxy luminosity and $B/T$ (uncorrelated for the ``B/T bias'' hypothesis, positive for the ``Bulge Enhancement'' hypothesis, and negative for the ``Disk Fading'' hypothesis). This approach is mathematically equivalent to the one we pursue here. We choose our approach because disk fading and bulge enhancement act independently on disk and bulge luminosities, so it is more straightforward to use the correlations of these quantities, rather than total luminosity, with $B/T$ to discriminate among the three null hypotheses.}

We compare the Monte Carlo samples generated by each of the three null hypotheses to the observations by fitting regression lines to the $M^{*}_{bulge}(\alpha=-1)$-$x$ and $M^{*}_{disk}(\alpha=-1)$-$x$ relations both in the observed and Monte Carlo samples, where $x=ln[(B/T)/(1-B/T)]$. The coordinate transformation from $B/T$ to $x$ ensures that the difference between the slopes of the relations does not depend on $B/T$, but it does not change the sense of the correlation. We perform the linear regression over a variety of ranges in $B/T$ and compare models and observations separately for each of them. This allows us to test whether a given null hypothesis might still describe a viable transformation mechanism over a smaller range of morphologies (e.g., from spirals to S0s), even if it cannot explain the observed trends over the full range $0\leq B/T\leq1$.

\begin{figure}
\plotone{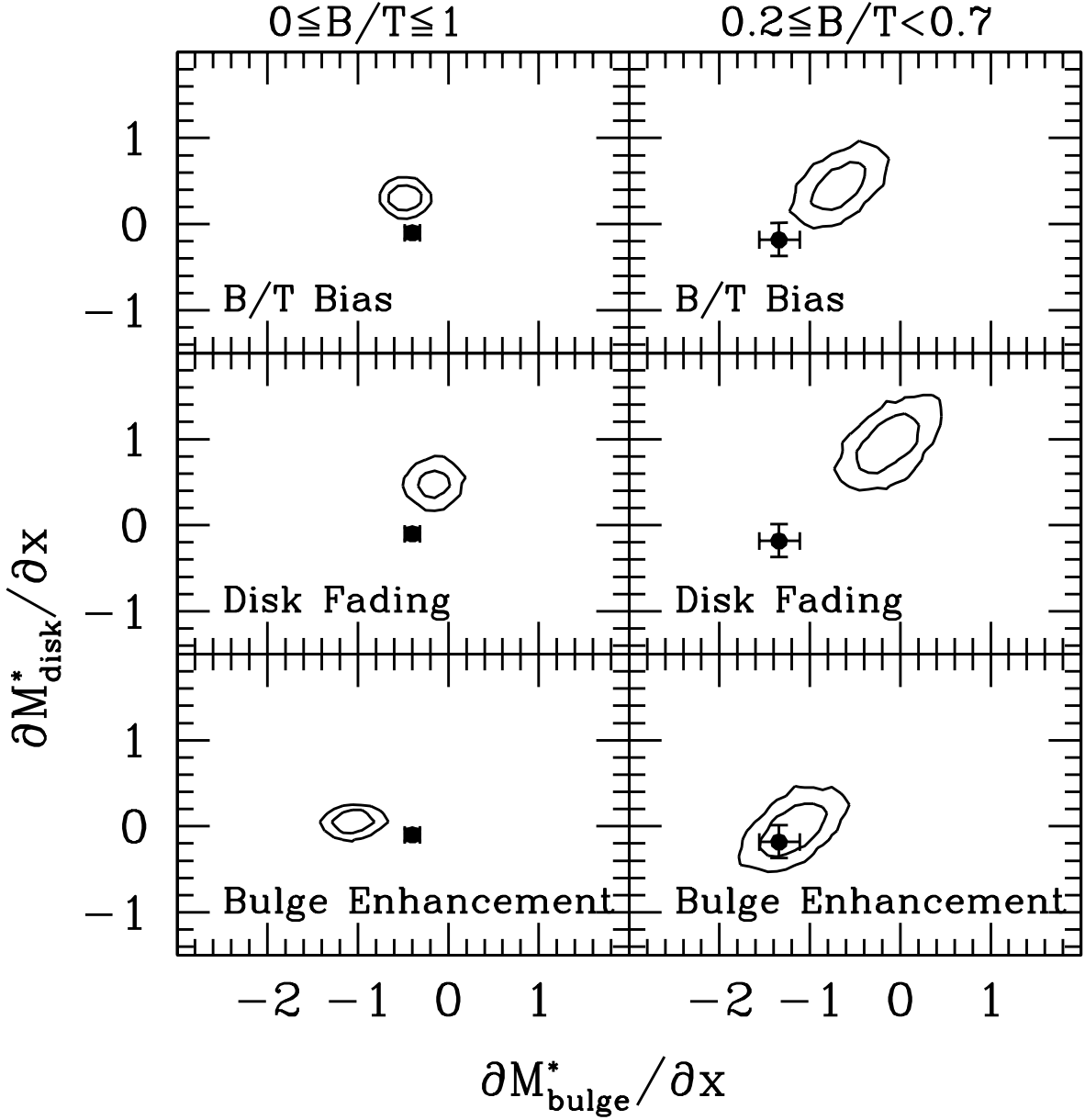}
\caption{Observed slopes for the $M^{*}_{bulge}(\alpha=-1)$-$x$ and $M^{*}_{disk}(\alpha=-1)$-$x$ correlations (filled circles), compared with 1- and 2$\sigma$ error contours from the three null hypotheses. The quantity $x$ is a parametrization of $B/T$. Left panels show fits over the full range of $0\leq B/T \leq 1$, right panels show fits over the range $0.2\leq B/T < 0.7$, which extends from early-type spirals to S0s and ellipticals.  No single null hypothesis can explain the entire morphological sequence. Over the range $0.2\leq B/T < 0.7$, bulge enhancement is clearly favored.}
\label{f3}
\end{figure}

Fig. \ref{f3} shows the results of these comparisons for all three null hypotheses over two different ranges in $B/T$. Plotted as contour lines in ($\partial M^{*}_{bulge}(\alpha=-1)/ \partial x$,$\partial M^{*}_{disk}(\alpha=-1)/ \partial x$) space is the distribution of regression slopes recovered from the null hypotheses. The contour lines in each panel encompass the results from 95\% and 68\%, respectively, of the Monte Carlo runs for a given null hypothesis. The observed slopes are marked by a data point in each panel. If the observed slopes fall outside the 95\% contour line, we consider the observations inconsistent with the null hypothesis for that panel.

%f2.eps > f5.eps

We find the following results:

(1) None of our three null hypotheses alone, including B/T bias, can explain the observed $M^{*}(\alpha=-1)$-$x$ correlations over the entire morphological sequence. The left panels of Fig. \ref{f3} show the observed slopes relative to the distribution expected from each of the three null hypotheses. Excluding the $B/T<0.2$ subsample (which is not affected by the disk fading or bulge enhancement prescriptions anyway because it lies below the progenitor cutoff value of $B/T=0.3$) does not alter this conclusion. 

(2) The ``B/T Bias'' and ``Disk Fading'' hypotheses are clearly ruled out over the range $0.2\leq B/T < 0.7$, which excludes the most extreme late-type spirals and early-type ellipticals. In contrast, the ``Bulge Enhancement'' hypothesis is in excellent agreement with the observed correlations. The right panels of Fig. \ref{f3} show the observed slopes over this range, compared to the expected distributions.   

(3) If we tighten the range further to $0.2\leq B/T < 0.5$, the ``Disk Fading'' hypothesis is still ruled out. This range of $B/T$ covers the range between the peaks of the distribution of spirals and S0s, and thus provides the most specific constraints on the possible morphological transformation mechanisms affecting this sample. 

Thus, our Monte Carlo approach does more than account for $B/T$ bias and establish the presence of a physical effect --- it also demonstrates that even a very simplistic bulge enhancement model reproduces not only the sense, but also the magnitude of the observed $M_{disk}^{*}$-$B/T$ and $M_{bulge}^{*}$-$B/T$ relations over all $B/T$.

We emphasize that our analysis does not assume that early types evolve from a progenitor population like the late types in our sample. We only assume that both early and late types evolved from the same progenitor population in the past, prior to any subsequent environmental transformations. Relaxing this assumption might make it possible to reproduce the observed $M_{disk}^{*}$-$B/T$ and $M_{bulge}^{*}$-$B/T$ relations via disk fading over a limited $B/T$ range, but it is hard to imagine a model that would generate the observed correlations over all $B/T$.

\subsection{Bulge- and Disk Luminosity as a Function of Environment}
\label{bdenv}

The results of the previous section suggest that the morphologies of cluster galaxies are differentiated by bulge enhancement, not by disk fading. Is the morphology-environment relation in clusters \citep{dressler80} thus due mostly to changes in bulge, instead of disk, luminosity? In other words, is there evidence that the bulges of galaxies grow more luminous towards the denser centers of clusters? If disk fading does occur as a galaxy falls through a cluster, we should also see differences in disk luminosity with projected radial distance from the cluster center. 

We test whether the morphology-environment relation is due to bulge enhancement or disk fading by splitting the sample, in analogy to the analysis in \S \ref{bdlfs}, into six subsamples selected by projected radial distance from the cluster center. We then calculate $M^{*}_{bulge}(\alpha=-1)$ and $M^{*}_{disk}(\alpha=-1)$ for each subsample. To account for the fact that the six clusters in our sample have different intrinsic physical radii, we scale the radial distances by 800 km s$^{-1}$ $\sigma^{-1}$, where $\sigma$ is the line-of-sight cluster velocity dispersion. We refer to this scaled distance as $R_{\sigma}$. This scaling is motivated by the fact that the size of virialized systems increases approximately as $\sigma$ \citep{girardi}. The constant of 800 km s$^{-1}$ is typical of rich clusters \citep{zab90}.

 In this analysis, the uncertainties in the Schechter fits are typically larger than those found in \S \ref{bdlfs}, and the effect is too subtle to determine whether there is a significant correlation of $M^{*}_{bulge}$ or $M^{*}_{disk}$ with radius. Nonetheless, we obtain constraints on the slopes of the $M^{*}_{bulge}$-$R_{\sigma}$ and $M^{*}_{disk}$-$R_{\sigma}$ relations of $\partial M^{*}_{bulge}/\partial R_{\sigma}=(0.31\pm0.27)$ mag Mpc$^{-1}$ and $\partial M^{*}_{disk}/\partial R_{\sigma}=-0.02\pm0.20$ mag Mpc$^{-1}$.

Does a more sensitive test allow us to establish whether the $M^{*}_{bulge}$-$R_{\sigma}$ relation is significantly different from zero? We use a Spearman Rank Correlation test \citep{gibbons} to examine correlations between the bulge luminosities of individual galaxies (with a total absolute magnitude $M_{total}\leq-19.25$) and the radial distance $R_{\sigma}$. A rank correlation test is appropriate in this case, because the distributions of bulge and disk luminosities are non-linear. In calculating the coefficients, we weight each data point by the weighting factor assigned to the galaxy by the DML solution. To additionally improve the significance of our conclusions, we exclude A1060, because of its small sampling radius ($R_{\sigma}\approx 0.7^{-1}$ Mpc), and analyze galaxies only within $R_{\sigma}=1.8^{-1}$ Mpc, the largest radius to which the other five clusters are fully sampled. 

% m:   0.0431497352  0.309399735  0.575649735
% m:  -0.21916791 -0.0204179101  0.17833209

The rank correlation coefficient between bulge luminosity and projected radius for 937 galaxies is $r_{M_{bulge},R_{\sigma}}=0.16$, which is significant at the $5\sigma$ level. The correlation is weak because, for any given radius, galaxies have a wide range of absolute magnitudes. The effect is nonetheless significant, implying that bulges are systematically brighter at smaller projected radii. 

After performing a similar analysis on the disk luminosities, we find no correlation between disk luminosity and radius; the correlation coefficient is $r_{M_{disk},R_{\sigma}}=-0.02$, a deviation of only $0.5\sigma$ from zero. Does this result imply that the disk luminosities of individual galaxies remain constant as a function of radial distance? On its own, the fact that the correlation coefficient is close to zero does not permit such a conclusion, because effects at the bright and at the faint end of the disk luminosity function could cancel each other. For example, a simultaneous brightening of the brightest $M_{disk}$ and increase in the number of faint disks could conspire to keep the median disk luminosity constant. Likewise, the fact that $M^{*}_{disk}$ itself does not vary significantly with radius could also be explained if disk galaxies (which are likely to dominate the bright end of the disk LF) lie far from the cluster center in three-dimensional space and are only seen close to the core in projection. However, if we consider both observations together --- that neither the characteristic bright-end disk magnitude nor the median disk luminosity of all galaxies with $M_{R}<-19.25$ vary significantly as a function of projected radius --- then we can conclude that radial distance has no significant effect on the luminosities of individual galaxy disks.

The observed correlation of $M_{bulge}$ with $R_{\sigma}$, in which bulges are brighter closer to the projected cluster center, may at least partially explain the morphology-environment relation. This result is consistent with those in \S \ref{bdlfs}, and thus provides additional support for the hypothesis that the morphology-environment relation is affected by galaxy-galaxy interactions that enhance bulge luminosities, rather than by the disruption of star formation with subsequent passive evolution and fading of the disks.

\section{CONCLUSIONS}

Using a new survey of six clusters of galaxies \citep{cz2003}, we constrain the mechanisms by which early-type galaxies may evolve from late-type galaxies in two ways: 1) by comparing the luminosities of bulges and disks as functions of overall galaxy morphology, as measured by the bulge fraction $B/T$, and 2) by comparing the trends of bulge and disk luminosity with radial distance from the cluster center.

 Models that generate early type galaxies by fading the disks of late-type galaxies are ruled out for a wide range of morphologies. Specifically, the bulges of galaxies with $B/T\approx0.45$, typical of S0 galaxies, are significantly brighter than expected if disk fading were the primary mechanism transforming late-type into early-type galaxies. Bulge enhancement models, which increase $B/T$ by increasing the luminosity of the bulge, are in excellent agreement with our observations over a wide range of $B/T$. 

This result is strengthened by our comparison of bulge and disk luminosities as a function of projected radial distance from the cluster center. There is a significant tendency for bulges to be brighter towards the cluster center, while galaxy disks have a similar luminosity regardless of their position within the cluster. This test reveals no evidence that cluster-specific processes reduce the luminosity of disks, and thus does not support disk fading mechanisms such as ram pressure stripping or strangulation as shapers of the morphological sequence in clusters. Instead, our results favor processes like galaxy-galaxy interactions and mergers, which can enhance galaxy bulges. Such processes are more efficient in lower-density environments such as poor groups of galaxies, which, when accreted hierarchically by clusters, may have played a major role in generating the morphological sequence that we observe in rich clusters today.

\acknowledgements

We would like to thank Luc Simard, author of GIM2D, for helpful advice, and Jim Gunn and Dennis Zaritsky for valuable discussions. We are also grateful to the anonymous referee who provided valuable suggestions.

This research has made use of the NASA/IPAC Extragalactic Database (NED) which is operated by the Jet Propulsion Laboratory, California Institute of Technology, under contract with the National Aeronautics and Space Administration. 

AIZ and DC acknowledge support from NSF grant \# AST-0206084.

\end{document}